\documentstyle[12pt,epsfig]{article}                                                            
                                                            
\newlength{\dinwidth}                                                            
\newlength{\dinmargin}                                                            
\def\lapproxeq{\lower .7ex\hbox{$\;\stackrel{\textstyle                                                            
<}{\sim}\;$}}                                                           
\def\gapproxeq{\lower .7ex\hbox{$\;\stackrel{\textstyle                                                            
>}{\sim}\;$}}  
                                                                                          
\def\be{\begin{equation}}                                                            
\def\ee{\end{equation}}                                                            
\def\bea{\begin{eqnarray}}                                                            
\def\eea{\end{eqnarray}}                                                            
                                                            
\def\funp{{I\!\!P}}                           

\begin{document}                                                            
\titlepage                                                           
\begin{flushright}                                                            
IPPP/01/16 \\     
DCPT/01/32 \\                                                           
20 April 2001 \\                                                            
\end{flushright}                                                            
                                                           
\vspace*{2cm}                                                           
                                                           
\begin{center} 
{\Large \bf High $p_T$ Higgs signal for the LHC}\\  
                                                           
\vspace*{1cm}                                                           
V.A. Khoze$^a$, A.D. Martin$^a$ and M.G. Ryskin$^{a,b}$ \\                                                            
                                                          
\vspace*{0.5cm}                                                            
$^a$ Department of Physics and Institute for Particle Physics Phenomenology, University of     
Durham, Durham, DH1 3LE \\        
$^b$ Petersburg Nuclear Physics Institute, Gatchina, St.~Petersburg, 188300, Russia                   
\end{center}                                                           
                                                           
\vspace*{2cm}                                                           
                                                           
\begin{abstract}                                                            
We show that the broad transverse momentum distribution of the Higgs boson produced by 
$WW$ fusion can provide a viable way to identify $H \rightarrow 
b\bar{b}$ decays at the LHC, if particular kinematical configurations with large rapidity 
gaps are selected.  We estimate the 
event rate of the signal and of the QCD $b\bar{b}$ background.  We also discuss Higgs 
boson detection via the $H \rightarrow \tau\tau$ and $H \rightarrow WW^*$ decay modes.
\end{abstract}                                                 
            
\newpage                  
\section{Introduction}

One of the main problems of searching for an intermediate mass Higgs boson at a hadronic 
collider is that it is hard to observe the dominant $H \rightarrow b\bar{b}$ decay mode due to 
the huge QCD $b\bar{b}$ background.  An attractive possibility is to search for the 
process in which the Higgs boson is produced with a large rapidity gap on either side.  The 
cleanest situation is double-diffractive exclusive production
\be
\label{eq:a1}
pp \; \rightarrow \; p \: + \: H \: + \: p,
\ee
where the plus sign is used to indicate a rapidity gap (and similarly for $p\bar{p}$ collisions).  
However the predicted cross section is 
rather small \cite{KMR,KMRmm}\footnote{Note that the existing literature shows a wide 
range of predictions for this cross section which vary by many orders of magnitude.  These 
can be found, for example, in \cite{BL,KMRH97,KL,AR}.  It is worthwhile to mention that 
the recent estimates of Refs.~\cite{KMR,KMRmm} give the lowest cross section
 among those listed in \cite{AR}.}.  First, 
due to the proton form factors, the available phase space is strongly limited in the transverse 
momentum of the produced Higgs, $q_T \sim 1/R_p$ where $R_p$ is the radius of the 
proton.  Second, we must include the probability that the rapidity gaps survive the soft 
rescattering effects of spectator partons which may populate the gaps with secondary particles 
see, for instance, Ref.~\cite{BJ}.  Third, the cross section is also suppressed by QCD 
radiative effects.  That is by Sudakov-like suppression factors which allow for the possibility not to 
bremsstrahlung gluons which again may populate the rapidity gaps.

To enlarge the cross section we can consider semi-inclusive configurations \cite{KMRH97} 
in which the protons may dissociate,
\be
\label{eq:a2}
pp \; \rightarrow \; X \: + \: H \: + \: Y,
\ee
but where the Higgs is still isolated by rapidity gaps.  In this case there is no 
proton-form-factor suppression and the Higgs bosons populate a much larger $q_T$ phase 
space.  Simultaneously the QCD radiative suppression becomes weaker, since the Sudakov 
double $\log$ takes the form $\sim \alpha_S \ln^2 (M_H/\langle q_T\rangle)$, where now 
$\langle q_T \rangle \gg 1/R_p$.  Moreover a significant contribution to process (\ref{eq:a2}) 
comes from Higgs production via $WW$ fusion (see Fig.~1(a)), where on account of the large $W$ boson 
mass the cross section is rather flat in $q_T$.  Furthermore, since this process 
is mediated by $t$-channel $W$ exchange, which is a point-like colourless object, there is no 
corresponding bremsstrahlung in the central region \cite{DKS} and thus the Sudakov 
suppression of the rapidity gaps does not occur. 

\begin{figure}[htb]  
\begin{center}\
\epsfig{figure=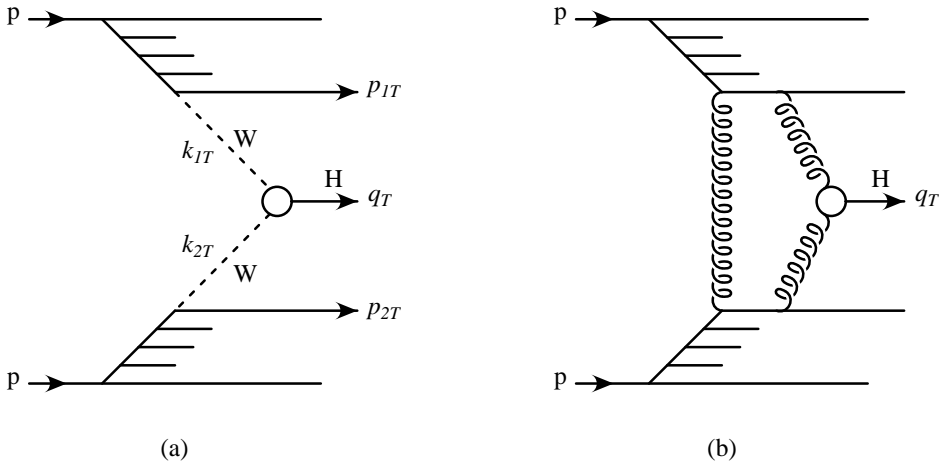,height=2.5in}  
\caption{Semi-inclusive Higgs production, $pp \rightarrow X + H + Y$, via (a) $WW$ 
fusion and (b) Pomeron-Pomeron fusion, where the QCD Pomeron is described by two-gluon 
exchange.
\label{fig:fig1}}  
\end{center}
\end{figure}

Another contribution to (\ref{eq:a2}) comes from the QCD subprocess $gg \rightarrow H$, 
where the colour flow of the hard $t$-channel gluons is screened by an accompanying, 
relatively soft, $t$-channel gluon as in Fig.~1(b).  The double-gluon-exchange mechanism 
was first discussed in Ref.~\cite{BL} (see also, for example, \cite{PP}) in terms of Higgs 
production by Pomeron-Pomeron fusion, using a non-perturbative two-gluon model of the 
Pomeron.  However it was shown \cite{KMR} that, at best, the $\funp\funp \rightarrow H$ 
mechanism gives a contribution comparable to $WW \rightarrow H$, for large rapidity gaps.  
On the other hand, $\funp$ fusion is the dominant mechanism for the QCD $b\bar{b}$ 
background to the semi-inclusive $H \rightarrow b\bar{b}$ production process (\ref{eq:a2}).  
In this respect the exclusive process (\ref{eq:a1}) appears to offer a better signal/background 
ratio since $\funp\funp \rightarrow q\bar{q}$ vanishes as $m_q/E_T \rightarrow 0$ in the 
forward direction due to a specific $J_z = 0$ selection rule \cite{P,KRMjj,KMRmm}, and 
$q\bar{q}g$ production is suppressed.  $E_T$ is the transverse energy of one of the jets.  As 
mentioned above, the only problem is that the predicted cross section is too small to exploit 
the exclusive Higgs signal, at least at the Tevatron.

For semi-inclusive production there is no $J_z = 0$ selection rule to suppress the $b\bar{b}$ 
background.  Moreover the expected $b\bar{b}$ mass resolution is worse than in the 
exclusive case.  Thus the signal-to-background ratio is relatively small \cite{KMR01}, 
\be
\label{eq:b2}
\frac{S}{B} \; \sim \; 0.01 \: \left ( \frac{M_H}{100~{\rm GeV}} \right )^3 \; \left ( 
\frac{4~{\rm GeV}}{\Delta M} \right ),
\ee
where $\Delta M$ is the mass resolution.  Nevertheless, we will show that it is possible to select 
a kinematic domain where semi-inclusive Higgs production may be identified at the LHC.  We exploit 
the much flatter $q_T$ dependence of semi-inclusive production and select Higgs candidates with large 
$q_T$, say $q_T > q_0$.  We show that it is possible to tune the $q_T$ and the rapidity cuts 
to select a domain where the predicted cross section is not too small so that the Higgs stands 
out from the background\footnote{The idea to increase the signal-to-background ratio by selecting 
high $q_T$ Higgs inclusively produced by $WW$ fusion, and to suppress the $q\bar{q} \rightarrow ZZ$ 
background which is steeper in $q_T$, was originally proposed in \cite{CEKS}.}.  We use the formalism of 
Ref.~\cite{KMR}, and include the recent 
evaluations of the survival probabilities of the rapidity gaps \cite{KMRsoft,KKMR}.  The 
calculation of the $H \rightarrow b\bar{b}$ signal is described in Section~2, 
and the computation of the QCD $b\bar{b}$ background is the subject of Section~3.  
Numerical predictions for the signal and background are given in Section~4 for particular 
choices of the large $q_T > q_0$ cut and of the rapidity gaps.

In Section~5 we discuss the possibility of observing the Higgs boson via process (\ref{eq:a2}) 
in the $H \rightarrow \tau^+ \tau^-$ decay mode or, as the Higgs becomes heavier, by $H \rightarrow 
WW^*$ decays.  In both of these cases the branching ratio for an intermediate mass Higgs is 
much smaller than that for $H \rightarrow b\bar{b}$, but there is almost no QCD 
background, provided that we select events with rapidity gaps.  In Section~6 we give our 
conclusions.

\section{The $WW \rightarrow H \rightarrow b\bar{b}$ signal at large $q_T$}

The cross section for electroweak Higgs production of Fig.~1(a) is well known 
\cite{GHKD,SPIRA}.  To obtain the $q_T$ distribution of the Higgs we need to perform the 
integration \cite{CEKS}
\be
\label{eq:a3}
\int \: \frac{d^2 k_{1T} d^2 k_{2T}}{(k_{1T}^2 + M_W^2) (k_{2T}^2 + M_W^2)} \; 
\delta^{(2)} \: (\mbox{\boldmath $k$}_{1T} + \mbox{\boldmath $k$}_{2T} - q_T) \; \ldots ,
\ee
where $\mbox{\boldmath $k$}_{1T,2T}$ are the transverse momenta of the exchanged 
$W^\pm$ bosons.  The parton-parton luminosity, which controls the normalisation of the 
cross section, was calculated using MRST partons \cite{MRST}.  At first sight it appears 
sufficient to evaluate the parton distributions at scales $k_{iT}^2$, and at the corresponding 
$x$ values, but the situation is not so trivial.  The problem is that the partons coupled to the 
$W$ bosons are emitted with rather large transverse momenta, $p_{1T}$ and $p_{2T}$, and 
materialise as jets with secondaries which may lie inside the rapidity gaps.  In order not to 
have jets with rapidity close to that of the Higgs boson, that is to have $| \eta_{\rm jet}| > 
|\eta_{\rm min}|$, we have to sample partons with light cone momentum fractions $x_i > 
x_{\rm min}^i$, with 
\be
\label{eq:a4}
x_{\rm min}^i \; = \; (M_H + p_{iT} \: \exp (|\eta_{\rm min}^i|)/\sqrt{s},
\ee
see also \cite{KRMjj}.  Here we have assumed that the Higgs boson is produced with 
rapidity\footnote{For $\eta_H \neq 0$, we simply make a boost and multiply the right-hand-
side of (\ref{eq:a4}) by $\exp (\eta_H)$.} $\eta_H = 0$.

\begin{figure}[htb]  
\begin{center}\
\epsfig{figure=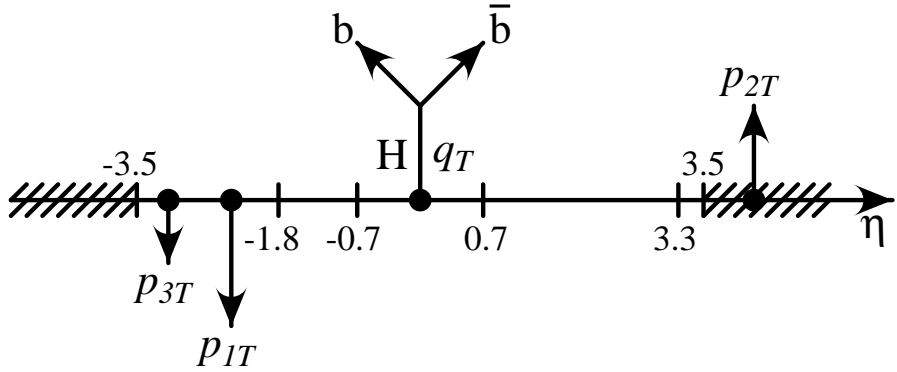,height=2in}  
\caption{The particular configuration of the rapidity gaps of the process of Fig.~1(a) 
used to calculate the cross sections given in Table~1.  We also include the configuration in 
which the diagram is reflected in the origin.  We show the configuration for $\eta_H = 0$, but 
we allow $\eta_H \neq 0$ and apply an overall Lorentz boost accordingly.
\label{fig:fig2}}  
\end{center}
\end{figure}

In order to retain a large part of the cross section, and also to have a favourable 
signal-to-background ratio, the experimental cuts must be chosen with care.  For illustration 
we calculate the event rates for the particular configuration shown in Fig.~2.  
 We require that the jet, say jet 2, with the smaller $p_T$ ($p_{2T} < p_{1T}$) satisfies  
\be
\label{eq:a6} 
\Delta \eta_2 \; = \; \eta_2 \: - \: \eta_H \: > \: 3.3,
\ee                                                                                  
while we allow jet 1 with the largest
 $p_T$ to be  possibly closer to the Higgs 
\be
\label{eq:a5}
\Delta \eta_1 \; = \; \eta_H \: - \: \eta_1 \: > \: 1.8.
\ee
Thus we have a rapidity gap $\Delta \eta > 5.1$, except for the $H \rightarrow b\bar{b}$ 
decay.  Moreover, within the overall rapidity interval $| \eta | < 3.5$ we require no other 
jets, apart from the $b,\bar{b}$ jets and possibly the two jets coupled to the exchanged bosons.  
However we allow for the possibility of one extra jet arising from the usual parton structure 
function evolution associated with the larger $p_T$ jet lying in the interval with the smaller $| 
\eta_{\rm min}|$, see Fig.~2.  In the leading $\log$ approximation the separation between 
these two jets (denoted $p_{1T}$ and $p_{3T}$ on Fig.~2) should be $\Delta \eta \gg 1$, but 
in reality the expectation is $\Delta \eta \sim 2$.  We emphasize that the $p_{iT}$ jets do not 
have to be within the rapidity interval $| \eta | < 3.5$.  The requirement is that the $p_{1T}$ and 
$p_{3T}$ jets have $\eta < -1.8$, and the $p_{2T}$ jet has $\eta > 3.3$.  The configuration 
reflected in the origin ($\eta = 0$) is also allowed.  This combination of rapidity gaps and jets 
(together with the possible tagging of $b$-jets and the reconstruction of their vertices) can 
provide a strong signature for Higgs production.  The predicted cross sections corresponding 
to the configurations allowed by Fig.~2 are presented in Section~4.

\section{The QCD $b\bar{b}$ background}

The $b\bar{b}$ background is calculated using the formalism described in  
Ref.~\cite{KRMjj} for the same jet configurations as given above.  The cross section is given 
by the convolution of the parton-parton luminosity and the production of $b\bar{b}$ in a 
colour-singlet configuration via the fusion of two BFKL Pomerons, see Fig.~3. The 
$\funp\funp \rightarrow b\bar{b}$ part of the cross section is given by
\be
\label{eq:a7}
\frac{d \sigma}{d E_T^2 d \eta_{b\bar{b}} d \Delta \eta} \; = \; \alpha_S^4 \: \frac{81}{64 
\pi^2} \: {\cal I} \left [ \frac{\pi \alpha_S^2 (E_T^2)}{6 E_T^2 M_{b\bar{b}}^2} \; \left (1 \: 
- \: \frac{2 E_T^2}{M_{b\bar{b}}^2} \right ) \right ],
\ee
where $\Delta \eta = | \eta_b - \eta_{\bar{b}} |$ and the expression in brackets is the $gg 
\rightarrow b\bar{b}$ colour-singlet hard subprocess cross section $d \hat{\sigma}/d \hat{t}$ 
\cite{P,KRMjj}.  The QCD Pomerons, each represented by two-gluon exchange, are 
described by BFKL non-forward amplitudes \cite{KRMjj,FR}.  Non-forward because the 
dominant 
contribution comes from the asymmetric region where the transverse momentum $Q_T$ 
carried by the screening gluon is much smaller than the total momentum transfer carried by 
the Pomeron.  Due to the asymmetry we have, besides $\Delta \eta_i$, a second logarithm, 
$\ln (k_{iT}^2/Q_T^2)$ in the BFKL evolution.  The summation of the double logarithms 
accounts for the probability not to emit extra gluons within the rapidity gap covered by the 
Pomeron.  In addition, we must include the usual Sudakov form factors which arise from the 
requirement that there is no gluon emission in the intervals $k_{iT}$ to $E_T$.  The factor 
${\cal I}$ in the cross section formula (\ref{eq:a7}) arises from the integration over the 
$t$-channel gluon loop in the amplitude of the process shown in Fig.~3.  ${\cal I}$ contains 
the BFKL amplitude and all the suppression factors arising from the requirement that there 
should be no gluon emission in the rapidity gaps, and it is given by eq.~(27) of Ref.~\cite{KRMjj}.

\begin{figure}[htb]  
\begin{center}\
\epsfig{figure=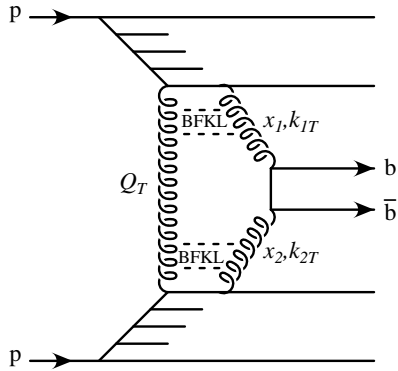,height=2in}  
\caption{The QCD $b\bar{b}$ background process to the $H \rightarrow b\bar{b}$ 
signal of Fig.~1(a).
\label{fig:fig3}}  
\end{center}
\end{figure}

Of course the $gg \rightarrow b\bar{b}$ cross section, $d \hat{\sigma}/d \hat{t}$, becomes 
too large at small $E_T$, see (\ref{eq:a7}).  Thus, in order to suppress this QCD $b\bar{b}$ 
background we impose a cut $E_T > 50$~GeV in the $b\bar{b}$ rest frame.  For a dijet 
system of mass $M_{b\bar{b}} = 115$~GeV this corresponds to the restriction that $b$ and 
$\bar{b}$ jets have polar angles $\theta > 60^\circ$.  The same cuts must be applied to the $H 
\rightarrow b\bar{b}$ decay and as a consequence we lose about half of the signal.  In terms 
of rapidities it means that we select events with jets with $\Delta \eta = | \eta_b - 
\eta_{\bar{b}} | < 1.4$, see Fig.~2.

\section{Predicted rates for $H \rightarrow b\bar{b}$ and background}

We have integrated the cross sections described above over the $b\bar{b}$ transverse 
momentum interval $q_T > q_0$ (with $q_0 = 25$ or 40~GeV), and over the rapidity of the 
$b\bar{b}$ pair.  The main contribution comes from the central region, $| \eta_{b\bar{b}} | < 
1.5$.  We use a fixed\footnote{In fact, using running $\alpha_S$ in the double logarithmic 
form of the BFKL non-forward amplitude one obtains, after the loop integration over $Q_T$, 
essentially the same result $(\sim \alpha_S (k_{iT}^2)/\Delta \eta)$ as for the case of fixed  
$\alpha_S$.  Since we select $b\bar{b}$ events with $q_T > 25$~GeV, the transverse momentum of 
the harder gluon is in the region 10-30~GeV, corresponding to $\alpha_S = 0.17$.} coupling 
$\alpha_S = 0.17$, which represents the typical coupling in the selected kinematical domain, 
see also Ref.~\cite{CFL}.

\begin{table}[htb]
\begin{center}
\begin{tabular}{|c|c||c|c|c||c|c|} \hline
& & Total & Signal & Background & & \\
$M_H$ & $q_T >$ & $\sigma_H$ & $\sigma_{H \rightarrow b\bar{b}}$ & 
$\sigma_{\funp\funp \rightarrow b\bar{b}}$ & $\sigma_{H \rightarrow \tau\tau}$ & 
$\sigma_{H \rightarrow WW^*}$ \\ \hline\hline
115 & 25 & 38 & 9.6 & 142 & 0.27 (0.54) & 3.1 \\
& 40 & 21 &5.3 &38 & 0.15 & 1.7 \\ \hline
140 & 25 & 29 & 3.3 & 61 & 0.09 (0.19) & 16 \\
& 40 & 14 & 1.6 & 20 & 0.04 & 7.6 \\ \hline\hline
115 & 25 & 61 & 16 & 116 & 0.44 (0.82) & 5.0 \\
& 40 & 35 & 9 & 36 & 0.25 & 2.8 \\ \hline
140 & 25 & 48 & 5.4 & 51 & 0.15 (0.29) & 27 \\
& 40 & 24 & 2.7 & 19 & 0.08 & 13 \\ \hline
\end{tabular}
\caption{$M_H$ and $q_T$ are in GeV.  The cross sections are in fb, and correspond to the 
rapidity cuts shown in Fig.~2, {\it except} for the $H \rightarrow \tau\tau$ values shown in 
brackets which correspond to the softer cuts given in the text.  Unlike the total $\sigma_H$, 
the $H \rightarrow b\bar{b}$ signal and background cross sections include the $H \rightarrow 
b\bar{b}$ branching fraction, the $b$ and $\bar{b}$ tagging efficiency and the polar angle 
$\theta > 60^\circ$ cut on the $b$ and $\bar{b}$ jets.  The upper and lower halves of the 
Table correspond to using the survival probabilities of the rapidity gaps that were determined 
in Refs.~\cite{KMRsoft} and \cite{KKMR} respectively.}
\end{center}
\end{table}

In Table~1 we present the signal and background rates for two different values of the Higgs 
boson mass and for the two choices of the $b\bar{b}~q_T$ cut.  The upper and lower halves 
of the Table correspond to different treatments of the survival probabilities of the rapidity 
gaps, as explained below.  The column of values of $\sigma_H$ shows the $WW \rightarrow 
H$ cross sections which allow for the rapidity gaps of Fig.~2 and for the $q_T > q_0$ cut, but 
which do {\it not} include the $H \rightarrow b\bar{b}$ branching ratio, or the $\theta > 
60^\circ$ jet cut or for the efficiency of the $b$ and $\bar{b}$ jet tagging.  If we include these 
latter effects\footnote{We assume a combined efficiency of 0.7 for identifying both $b$ and 
$\bar{b}$ jets.} then we have the \lq useful\rq\ $WW \rightarrow H \rightarrow 
b\bar{b}$ signal shown in the next column in Table~1, followed by the cross section for the 
QCD $\funp\funp \rightarrow b\bar{b}$ background.

These cross section values shown in the Table correspond to (\ref{eq:a7}) convoluted with 
the parton-parton luminosity, with (\ref{eq:a7}) integrated over the $b$ and $\bar{b}$ jet 
rapidities, $\eta_{b\bar{b}}$ and $\Delta \eta$, and over a small bin of transverse energy 
which corresponds to $b\bar{b}$ events in the Higgs mass interval.  The smallness of this 
interval is limited by the experimental jet resolution.  Here we assume $\Delta E_T = 4$~GeV 
in the $b\bar{b}$ centre-of-mass frame.

The predictions in the top half of the Table correspond to using the values of the survival 
probability $S^2$ listed in the double-diffractive (DD) column of Table~1 of 
Ref.~\cite{KMRsoft} for $\sqrt{s} = 14$~TeV.  That is $S^2 = 0.15$ for the $WW 
\rightarrow H$ signal (where we assume that the spatial distribution of the quarks is described 
by the electromagnetic form factor of the proton with slope 5.5~GeV$^{-2}$) and $S^2 = 
0.10$ for the $\funp\funp \rightarrow b\bar{b}$ background (where the slope of the 
corresponding distribution is taken to be 4~GeV$^{-2}$).  In this case for an integrated 
luminosity of 100~fb$^{-1}$ at the LHC\footnote{Of course, at large LHC luminosities secondary 
particles produced in \lq pile-up\rq\ events may fill the rapidity gaps.  However we hope 
that it is possible to select experimentally tracks coming from the same vertex and so separate 
the particles which belong to the event of interest, in which a Higgs boson is produced with a 
large rapidity gap on either side.} we have, for $M_H = 115$~GeV and $q_T > 
25$~GeV, about 1000 $WW \rightarrow H \rightarrow b\bar{b}$ identified events sitting on 
top of a QCD $b\bar{b}$ background of 14,000 events, see Table~1.  This would give an 8 
standard-deviation signal.  Increasing the $q_T$ cut improves the 
signal/background ratio, but decreases the number of events, so in fact the quality of the 
signal declines if, for example, we were to choose the cut $q_T > q_0$ with $q_0 = 50$~GeV.

The above values of the survival probability $S^2$ of the rapidity gaps were calculated 
\cite{KMRsoft} using a two-channel rescattering eikonal in which the diffractive 
eigen-channels have different cross sections of absorption, $\sigma_0 (1 \pm \gamma)$ with 
$\gamma = 0.4$.  In Ref.~\cite{KKMR}, arguments were given that the lower cross section 
arises mainly from the valence quark configurations and that the higher cross section comes 
dominantly from the gluon and sea quark configurations.  Adopting this simplified model 
would give a larger $S^2$ for $WW \rightarrow H$ production\footnote{In 
principle we can measure the survival probability $S^2$ for the gaps surrounding $WW 
\rightarrow H$ fusion by observing the closely related central production of a $Z$ boson with 
the same rapidity gap and jet signature \cite{CZ}.} where the valence quarks play a dominant 
role, and a lower $S^2$ for the QCD $\funp\funp 
\rightarrow b\bar{b}$ background, which originates from the gluons.  
 Of course, now the \lq survival' factor $S^2$ depends on the values of the mass and
$q_T$ of the Higgs boson (or $b\bar{b}$-pair). For smaller values of the mass and $q_T$ the 
screening corrections are stronger, since there is a larger contribution caused by gluon-gluon 
collisions.  For a pure gluon-gluon interaction the factor $S_{gg}^2=0.033$ in this
model, while for a valence quark collision $S^2_{qq}=0.37$ (and $S^2_{qg}=0.15$ for the 
case of gluon and valence quark collisions).

Averaging over all contributions we obtain a suppression factor $S^2\simeq 0.08
\, -\, 0.1$ for QCD $b\bar{b}$-pair production, whereas $S^2\simeq 0.24\, -\,
 0.26$ for Higgs production via the $WW$-fusion.  The limits of the range of $S^2$ correspond 
respectively to the largest and smallest values of $q_T$ and mass in Table 1.  As was
expected the factor $S^2$ is closer to $S_{gg}^2$ for the case of QCD
$b\bar{b}$ double-Pomeron production, but for Higgs production it is closer to
the $S^2_{qq}$ value. The results for this model are shown in the lower half of Table~1.  Thus for a 
luminosity 100~fb$^{-1},~M_H = 115$~GeV and $q_T > 25$~GeV, we have a chance to identify 1600 
$H \rightarrow b\bar{b}$ events sitting on a background of 11,600 events.  This would be about 
a 15 standard-deviation effect.
 To put it another way, a 
luminosity of  12~fb$^{-1}$ would be enough to achieve a 5 standard-deviation signal.

Of course, the above cuts and corresponding predictions are just examples.  The experimental 
cuts should be optimized, taking into account the specifics of the detectors.  Also note that 
there is a factor of two uncertainty in the background prediction due to the use of the double 
$\log$ approximation.  Fortunately the single $\log$ contributions are suppressed in our 
asymmetric two-gluon-exchange domain, so that we are not so sensitive to the uncertain 
higher-order BFKL effects.

\section{$WW \rightarrow H \rightarrow \tau^+ \tau^-$, $WW^*$ and $ZZ^*$ high $q_T$ 
Higgs signals}

Another possibility is to observe the $H \rightarrow \tau^+ \tau^-$ decay mode, where there is 
practically no QCD background.  Of course, the small $H \rightarrow \tau^+ \tau^-$ 
branching fraction leads to a small cross section, as shown in Table~1.  However we may 
increase the signal by choosing softer cuts.  For example, the values of the cross section 
shown in brackets correspond to the cut $q_T > 20$~GeV, and the rapidity cuts of the 
accompanying jets $| \eta_1 | > 1.5$ and $| \eta_2 > 2.9$ (for the case $\eta_H = 0$, as in 
Fig.~2).

The main background for the $H \rightarrow \tau^+ \tau^-$ signal comes from the central 
production of the $Z$ boson and its subsequent $\tau^+ \tau^-$ decay.  If we were able to 
reconstruct the mass of the $\tau^+ \tau^-$ pair it would be easy to identify $H \rightarrow 
\tau^+ \tau^-$ events.  Unfortunately there are two unobserved $\nu_\tau$ neutrinos from the 
$\tau$ decays.  It does not mean $M_{\tau\tau}$ is completely unknown.  It may be estimated 
from the decay configurations\footnote{For example, in the Higgs rest frame the $\tau^+$ and 
$\tau^-$ emerge back-to-back.  Since $M_H \gg m_\tau$, the direction of the decay products 
is, to a good approximation, collinear with the parent $\tau$. Hence we can find the Lorentz 
boost, $\mbox{\boldmath $\gamma$} = \mbox{\boldmath $q$}_H/M_H$, needed to restore 
the collinearity of the two $\tau$'s.  Also the transverse momentum $\mbox{\boldmath 
$q$}_T$ can be measured as the momentum balancing that of the jets or simply as the 
missing $E_T$.  Hence we can estimate the value of $M_H$.}, but the accuracy is not so 
good.  The cross section for the central production of a $Z$ boson, accompanied by two jets, 
has been calculated for the LHC energy in Ref.~\cite{DZ}, however without including the 
survival probability $S^2$ of the rapidity gaps.  If we include $S^2$ in their results then the 
$(Z \rightarrow \tau\tau) + 2$ jet cross section is predicted to be about 6~fb for the cuts 
similar to the ones that were chosen for the larger $H \rightarrow \tau^+ \tau^-$ signal shown 
in brackets in Table~1.  The background is therefore an order of magnitude, or more, larger 
than the Higgs signal.  Nevertheless, if the mass resolution is not too bad, there is a chance to 
identify the $H \rightarrow \tau^+ \tau^-$ signal.  Clearly the $Z \rightarrow \tau^+ \tau^-$ 
decay mode will pose less of a problem the higher the value of $M_H$.

For larger values of $M_H$ the $H \rightarrow \tau^+ \tau^-$ decay mode decreases as the 
$H \rightarrow WW^*$ decay opens up.  Therefore for the heavier Higgs boson it is more 
promising to search for the $H \rightarrow WW^*$ and $H \rightarrow ZZ^*$ signals\footnote{It was 
shown in Ref.~\cite{RZP} that the $H \rightarrow WW^* \rightarrow \ell^+ \ell^- +~{\rm missing}~p_T$ 
signal may be considered as a discovery mode, even for a light $(M_H = 115~{\rm GeV})$ Higgs boson, 
if one selects events where the Higgs is produced by $WW$ fusion in association with two light 
quark jets.  The corresponding kinematics (similar to that shown in Fig.~2) were discussed in 
detail in \cite{RZP}.  However another criteria for the large rapidity gap was used in \cite{RZP} 
(within the gap no jets with $p_T >20$~GeV are permitted) and so another gap-survival factor was 
used.}.  The corresponding cross sections are listed in the last column of Table~1.  We see, for 
$M_H = 140$~GeV, that the $H \rightarrow WW^*$ cross section is about 20~fb.  Again there is 
practically no QCD background in the configuration with two large rapidity gaps either side 
of the parent Higgs.  Of course, we must allow for the detection efficiency of the various 
decay modes.  It is difficult to extract the value of $M_H$ from the leptonic decays of both 
the $W$ and $W^*$.  However it may be possible to use the decay configuration $W 
\rightarrow$ two quark jets and $W^* \rightarrow \ell \nu$.  On the other hand, the $H 
\rightarrow ZZ^* \rightarrow$ 4 leptons process will provide a rather clean signature.

Finally, we emphasize if $M_H > 2~M_Z$, then adding the rapidity signature to the 
gold-plated $H \rightarrow ZZ \rightarrow$ 4 lepton channel, would practically eliminate the 
background due to $q\bar{q} \rightarrow ZZ$.  The Higgs signal may be thus purified at the 
expense of introducing the survival probability factor $S^2$, and in this way allow a more 
precise study of the properties of the Higgs boson.

\section*{Acknowledgements}

One of us (VAK) thanks the Leverhulme Trust for a Fellowship.  This work was partially 
supported by the UK Particle Physics and Astronomy Research Council,  by
the EU 
Framework TMR programme, contract FMRX-CT98-0194 (DG 12-MIHT) and by the
RFFI grants 00-15-96610 and 01-02-17095.

\end{document}